\documentclass[final,times,5p,twocolumn,fleqn]{elsarticle}

\usepackage{natbib}
\usepackage{graphicx}
\usepackage{amsmath}
\usepackage{amssymb}
\usepackage{amsfonts}
\usepackage[pdfpagemode=none, pdfstartview=FitH]{hyperref}
\usepackage{mathrsfs}
\usepackage{dcolumn}

\newcommand{\re}{\mathop{\mathrm{Re}}}
\journal{Physics Letters B}

\begin{document}

\begin{frontmatter}

\title{Low-$Q^2$ scaling behavior of the form-factor ratios for the $N\Delta(1232)$-transition}

\author[SFU]{G.~Vereshkov}
	\ead{gveresh@gmail.com}
	
\author[SFU]{N.~Volchanskiy\corref{corauth}}
	\ead{nikolay.volchanskiy@gmail.com}
	
\address[SFU]{%
Research Institute of Physics,
Southern Federal University,
344090, Rostov-na-Donu, Russia
}

\cortext[corauth]{Corresponding author}

\begin{abstract}
We address the issue of the scaling behavior of the nucleon electromagnetic form factors (FFs). We propose a consistent Lagrangian of the electromagnetic $N\Delta(1232)$-interaction possessing all the internal symmetries of the spin-$\frac32$ baryon resonance, point and gauge invariance. The point and gauge invariant FFs may exhibit quasi-scaling behavior---while the FFs do not reach asymptotic scaling domain, the ratios thereof do. The hypothesis of the quasi-scaling is in good agreement with the experimental data available at $Q^2 \geqslant 0.4 \text{ GeV}^2$.
\end{abstract}


\end{frontmatter}


\section{Introduction}

The high-$Q^2$ scaling laws are believed to be firm and well understood predictions of perturbative QCD (pQCD) in the physics of the nucleon form factors (FFs) \cite{1973Brodsky-Farrar, 1975Brodsky-Farrar, 1980PhRvD..22.2157L, 2000PhRvD..62k3001B, 2000stefanis}. At the experimentally accessible energies, however, the asymptotic scaling behavior of the FFs seems not to be apparent. For instance, the measurements of the electromagnetic ratio $R_\text{EM} (Q^2)$ of the $\Delta(1232)$ resonance \cite{fr-99, ka-01, r2008PDG, sp-05, ke-05, 2007Julia-Diaz, 2008stave, 2009aznauryan, 2009villano} indicate that it remains to be small and negative up to the momentum transfer $Q^2 = 7 \text{ GeV}^2$, which disagrees sharply with the pQCD scaling $R_\text{EM} (Q^2 \to +\infty) = +1$ \cite{ca-86, 1988carlson}. It is possible that the transition to the scaling may not occur up to extremely large momentum transfers $Q^2 \gg 20 \text{ GeV}^2$ \cite{1996PhRvD..53.6509B, br-06}. On the other hand, it is a well established experimental fact that the ratio of the elastic nucleon FFs does exhibit the QCD scaling behavior at $Q^2 \geqslant 0.5 \text{ GeV}^2$, while the Dirac and Pauli FFs in themselves do not \cite{2003Belitsky, br-04}. In what follows such $Q^2$-evolution---the scaling behavior of the ratios of the FFs, providing that the FFs do not reach asymptotic scaling regime---is referred to as a \textit{quasi-scaling} behavior of the FFs.

The basic question considered in this Letter is whether the quasi-scaling behavior is a universal feature of the nucleon electromagnetic FFs or a peculiarity that is specific to the elastic FFs. If the former option is correct, it will provide an important meeting ground between pQCD and experimental physics. By phenomenological analysis we find that the hypothesis of the quasi-scaling does not contradict the experimental data available in the first and second resonance regions. In this short Letter we confine ourselves to the case of the nucleon-to-$\Delta(1232)$ transition FFs, since these quantities are the most extensively measured nucleon FFs, except for the elastic ones. The quasi-scaling fits to the data on the $pN^*(1440)$, $pN^*(1520)$, and $pN^*(1535)$ transitions will be given in a separate paper.

To investigate the $Q^2$-evolution of the $N\Delta(1232)$ FFs, we treat the resonance $\Delta(1232)$ as an effective Rarita-Schwinger (RS) field $\Psi_\mu$ \cite{ra-sh} (Hereafter we omit the spinor indices.) However, effective field theory of the nucleon electromagnetic FFs involves a problem to be resolved, before the rate of the transition to pQCD behavior of the FFs can be assessed. The problem is to specify the effective baryon-meson vertex for the radiative $N\Delta(1232)$-transition. This is of crucial importance, since the vertex determines relations between experimentally measurable helicity amplitudes $\mathscr{A}(Q^2)$, $\mathscr{A} = A_{3/2},\, A_{1/2},\, S_{1/2}$---counterparts of the elastic Sachs FFs---and a set of the transition FFs $F_f(Q^2)$, $f=1,\,2,\,3$ that are free of kinematic factors,
\begin{align}\label{HAs through FFs}
	\mathscr{A}(Q^2)
	= \mathscr{A} \bigl[ Q^2,\, F_1(Q^2),\, F_2(Q^2),\, F_3(Q^2)\bigr].
\end{align}
The relations \eqref{HAs through FFs} comprise two types of $Q^2$-dependent functions. The functions of the first type are the transition FFs $F_f(Q^2)$ to be evaluated in QCD or phenomenological approaches. The second type is kinematic factors that are specified by defining the effective vertex of the $N\Delta$-interactions with photons and vector mesons. Thus, the physical meaning of the FFs can depend upon the definition of the vertex, since different relations \eqref{HAs through FFs}, if resolved in respect to $F_f(Q^2)$, may result in different regimes of the $Q^2$-evolution of the FFs.

It is important to note the following property of the relations between the elastic Sachs helicity FFs $G_{M,\,E}(Q^2)$ and the Dirac and Pauli FFs $F_{D,\,P}(Q^2)$ \cite{1950rosenbluth, 1960ernst}. At asymptotically high $Q^2$ the magnetic FF scales like the Dirac one, $G_M(Q^2) \sim F_D(Q^2)$, and the electric FF scales like the Pauli one, $G_E(Q^2) \sim Q^2 F_P(Q^2)$ \cite{2003Belitsky}. It is the property that gives the FFs $F_{D,\,P}(Q^2)$ distinct interpretations in terms of the underlying quark dynamics---the Dirac FF is the FF of the processes conserving quark helicities, while the Pauli FF is the FF for the processes with quark-helicity flips. This property is a precondition to the fact that correct high-$Q^2$ behavior of the Sachs FFs is achieved by simple power-law asymptotic evolution of the Dirac and Pauli FFs. This, in turn, enables us to formulate the simple scaling relation for the ratio of the Dirac and Pauli FFs.

To analyze the applicability of the quasi-scaling to the data on the $N\Delta(1232)$-transition, we need the decomposition of the effective $N\Delta$-vertex that fulfills the same requirements as in the elastic case. The decomposition should result in such relations \eqref{HAs through FFs} that allow the FFs $F_f(Q^2)$ to get interpretation in terms of the underlying quark dynamics. More precisely, at high $Q^2$ the asymptotic behaviors of the helicity amplitudes $A_{1/2}(Q^2)$, $A_{3/2}(Q^2)$, $S_{1/2}(Q)$ should be governed by different FFs. In this way, the FFs $F_f(Q^2)$ acquire the asymptotic statuses of the corresponding amplitudes---the FF of the processes without quark-helicity flips, with one helicity flip, and with flips of two quark helicities. Besides, this guarantees that the high-$Q^2$ behavior of the amplitudes predicted by pQCD \cite{ca-86, 1988carlson, 2004idilbi} is achieved by simple power-law evolution of the FFs,
\begin{align}\label{}
	F_f(Q^2) \sim \frac{1}{Q^{2p_f} \ell^{n_f}},
	\qquad
	\ell = \ln{\frac{Q^2}{\Lambda^2}}
\end{align}
for some particular values of the exponents $n_f$ and $p_f$.

The effective $N\Delta$-vertex, however, is not as easy to define as in the elastic case. So far, several attempts have been made to fix the vertex. The first and most popular one \cite{jo-sc,Devenish1976} is based upon the considerations of simplicity. It results, however, into a number of mathematical pathologies, i.e., such interactions invoke unphysical lower-spin states of the RS field, if it is off its mass shell \cite{ve-zw,johnson-sudarshan,1998pascalutsa}. Fortunately, this and other inconsistencies can be remedied in an elegant way, if we require the invariance of the vertex under gauge transformations of the RS field \cite{1998pascalutsa,pa-ti}. Thus any consistent decomposition should be gauge invariant. However, the most general GI decomposition \cite{pa-ti} allows for many possible structures. In this sense, although applying strict constraints on the form of the vertex as it is, the requirement of the gauge invariance still leaves much unphysical redundancy in the general decomposition of the $N\Delta(1232)$-vertex. This brings up the question, Can we impose on the gauge invariant $N\Delta(1232)$-vertex such additional symmetry conditions that will postulate it uniquely, give us a set of the FFs satisfying the requirements in the domain of high-$Q^2$, and do not contradict the experimental data? We should also note that there is an example of such a theory---the elastic $\gamma^*NN$-vertex \cite{1950rosenbluth, 1960ernst} is determined uniquely by the electromagnetic gauge symmetry\footnote{The first Dirac term comes from the covariant derivative in the kinetic term of the nucleon Lagrangian and the second Pauli term is the only gauge invariant expression involving just one field derivative.}.

In this Letter we propose the effective vertex of the electromagnetic $N\Delta(1232)$-interactions possessing all the internal symmetries of the free RS field. We find that these symmetries allow us to constrain the tensor-spinor structure of the vertex and to provide a new set of the transition FFs which is defined unambiguously by the requirements of the symmetry and locality. After the problem of constructing the Lagrangian is solved, the hypothesis of the quasi-scaling can be assessed. We observe that the transition FFs introduced here may exhibit the quasi-scaling behavior. This hypothesis is in good agreement ($\chi^2/\text{DOF} = 1.03$) with the existing data down to as low as $Q^2 \approx 0.4 \text{ GeV}^2$.

All possible decompositions of the $N\Delta(1232)$-vertex are equivalent on the mass shell of resonance up to a linear redefinition of the FFs $F_f(Q^2)$. It can be expected that the quasi-scaling is present in all the models. Thus, it is interesting to compare the quasi-scaling in the model developed here and in the popular ones discussed in the literature \cite{jo-sc,pa-07}. We find that the hypothesis of the quasi-scaling of the FFs ratios in the non--gauge-invariant model \cite{jo-sc,vereshkov:073007} is in much poorer agreement with the data. In the same time, the gauge invariant FFs of Ref. \cite{pa-07} can not exhibit the quasi-scaling, only some linear combinations thereof could.

The remainder of this Letter is organized as follows. In Sections \ref{sec:symmetries} and \ref{sec:Lagrangian} we develop a phenomenological vertex for the electromagnetic $N\Delta(1232)$-interactions. The Section \ref{sec:VMD} reviews the vector-meson--dominance model of Ref.~\cite{vereshkov:073007}. In Section \ref{sec:fits} we contrast our model with the experimental data on the $Q^2$-dependence of the $N\Delta(1232)$-transition and test the hypothesis of the quasi-scaling behavior of the transition FFs. In Section \ref{sec:nPI} the quasi-scaling behavior of two popular FF sets is discussed. Finally, Section \ref{sec:conclusion} is a summary of our main results and conclusions.


\section{\label{sec:symmetries}Internal symmetries of the $\Delta(1232)$}

Comprising redundant degrees of freedom (DsOF) as it is, the RS field possesses peculiar internal symmetries \cite{ra-sh, 2004pilling}. The symmetries are inextricably linked to the constraints that are imposed on the reducible RS field in order to eliminate spurious DsOF \cite{pa-ti}. The free-field constraints can be written in a manifestly covariant form as $\partial^\mu \Psi_\mu = 0 = \gamma^\mu \Psi_\mu$ \cite{ra-sh}. Such simple structure of constraints is generated by a one-parameter equivalent class of the free-field Lagrangians \cite{moldauer-case, 2004pilling}
\begin{align}\label{total Lagrangian}
	&\mathscr{L}_\text{ff}(\lambda) =
	  \bar\Psi^\mu \left( i \Gamma_{\mu\nu\lambda} \partial^\lambda
	                      - M \Gamma_{\mu\nu} \right) \Psi^\nu
	                      +\text{H.c.},
	\notag\\
	&\Gamma_{\mu\nu\lambda} =
		g_{\mu\nu} \gamma_\lambda
		- \lambda^* \gamma_\mu g_{\nu\lambda} - \lambda \gamma_\nu g_{\mu\lambda}
		+ \left( \frac32 \lvert \lambda \rvert^2 - \re\lambda + \frac12 \right) 			         \gamma_\mu \gamma_\lambda \gamma_\nu,
	\notag\\
	&\Gamma_{\mu\nu} =
		g_{\mu\nu} -\left( 3 \lvert \lambda \rvert^2 - 3 \re\lambda + 1 \right) \gamma_\mu \gamma_\nu,
\end{align}
where $\lambda \neq \frac12$ is a complex parameter.

The equivalent class \eqref{total Lagrangian} of the free-field Lagrangians (not a Lagrangian for any $\lambda$ \footnote{To avoid confusion, it should be stressed that throughout this Letter the point invariance is taken to mean an invariance under the global point transformations of the RS field \emph{without shifting the free-field parameter $\lambda$ simultaneously so as to compensate the changes in the Lagrangian} (cf. Refs. \cite{1971nath-etemadi, 1989benmerrouche}). Thus, the equivalence class of the free-field Lagrangians is point invariant, while a Lagrangian is not for any $\lambda$.}) is invariant under the point transformations of the RS field, $\Psi'_\mu = \Theta_{\mu\nu}^{(\lambda,\lambda')} \Psi^\nu$, $\Theta_{\mu\nu}^{(\lambda,\lambda')} =
g_{\mu\nu} + \frac{\lambda'-\lambda}{2\left( 2 \lambda - 1 \right)} \gamma_\mu \gamma_\nu$ \cite{2004pilling}. The point transformations form a nonunitary symmetry group and shift the value of the Lagrangian parameter from $\lambda$ to $\lambda'$. Besides, in the massless case, the Lagrangian is invariant under the local gauge transformations, $\Psi'_\mu = \Psi_\mu + \Theta_{\mu\nu}^{(\lambda,1)} \partial^\nu \theta(x)$ \cite{ra-sh, 2004pilling}. Both the point and gauge transformations act only on the lower-spin components of the reducible RS field. Since such mixing the lower-spin components has nothing to do with the physics of the baryon resonances, it is suggestive that the interacting RS field should possess all the same symmetries as the free field. However, in the presence of interactions, the symmetries of the free RS field can be broken, which modifies covariant constraints and may result in different pathologies \cite{ve-zw, johnson-sudarshan, 1998pascalutsa}. To prevent such inconsistencies is possible, if we require the invariance of the interaction Lagrangian under the gauge transformation of the RS field \cite{1998pascalutsa, pa-ti}. The gauge invariant (GI) interactions, however, modify the free-field constraints making them nonlinear in the field operators \cite{1998pascalutsa}. Besides, the general GI decomposition of the $NR$-Lagrangian contains a great variety of couplings, the necessary three being chosen at will \cite{1998pascalutsa, pa-ti}.

It is of interest to study a more simple and constrained case of the interactions described by the vector-spinor currents $J_\mu$ satisfying the conditions of $p$- and $\gamma$-transversality, $\partial^\mu J_\mu = 0 =\gamma^\mu J_\mu$. In other words, we suggest synthesizing the ideas expressed earlier by Peccei \cite{1968peccei, 1969peccei} and Pascalutsa \cite{1998pascalutsa, pa-ti}. We consider the interaction Lagrangians that are invariant under both the point and gauge transformations of the RS field. Such point and gauge invariant (PGI) interactions preserve the structure of the free-field covariant constraints $\partial^\mu \Psi_\mu = 0 = \gamma^\mu \Psi_\mu$ and, therefore, lead to the first-order Dirac-like field equations $\bigl( i \hat\partial - M \bigr) \Psi_\mu = J_\mu$. As a consequence, the PGI interactions involve correct number of DsOF. Besides, they do not introduce the off-shell parameters, nor arbitrary or fixed ones. (The problem of the off-shell freedom is reviewed in Refs.~\cite{1971nath-etemadi, 1989benmerrouche}.)


\section{\label{sec:Lagrangian}PGI $N\Delta$-Lagrangian. Helicity amplitudes}

The general GI $\gamma^* N\Delta$-Lagrangian is a sum of the invariants $\bar\Psi^{\mu\nu,\bar\alpha} \Gamma_{\mu\nu\lambda\sigma\bar\alpha\bar\beta} N^{,\bar\beta} V^{\lambda\sigma}$, where $V_{\mu\nu}$, $V = \gamma, \,\rho(770), \,\rho(1450), \,\dots$ is a photon or vector-meson field strength; $\Gamma_{\mu\nu\lambda\sigma\bar\alpha\bar\beta}$ is a tensor spinor coupling antisymmetric in the indices $\mu$, $\nu$ and $\lambda$, $\sigma$; $\bar\alpha$ and $\bar\beta$ are multi-indices, the number of the indices therein being 2 less than the differential order of the Lagrangian term. The point invariance constrains the coupling tensor spinors by the condition of $\gamma$-transversality, $\gamma^\mu \Gamma_{\mu\nu\lambda\sigma\bar\alpha\bar\beta} = 0$. This condition provides us with a classification of the FFs by the differential order of the Lagrangian. The classification turns out to be naturally consistent with pQCD-inspired classification in terms of the quark-gluon subprocesses.

In the first permitted differential order there is just one term $I_1 = \bar\Psi^{\mu\nu} \Gamma_{\mu\nu\lambda\sigma} N V^{\lambda\sigma}$ in the Lagrangian, with the tensor spinor coupling being specified as
\begin{align}\label{kernel}
	\Gamma_{\mu\nu\lambda\sigma}
	=
	\frac13 \biggl[ &e_{\mu\nu\lambda\sigma} +
	i \gamma_5 \left(g_{\mu\lambda} g_{\nu\sigma} - g_{\nu\lambda} g_{\mu\sigma} \right)
	\notag\\&
	-\frac12 \left( g_{\mu\lambda} \tilde\sigma_{\nu\sigma}
	                - g_{\nu\lambda} \tilde\sigma_{\mu\sigma}
	                - g_{\mu\sigma} \tilde\sigma_{\nu\lambda}
	                + g_{\nu\sigma} \tilde\sigma_{\mu\lambda} \right)
	\biggr].
\end{align}
Here $\tilde\sigma_{\mu\nu} = \frac12 e_{\mu\nu\eta\xi} \sigma^{\eta\xi}$ is dual to $\sigma_{\mu\nu} = \frac12 (\gamma_\mu \gamma_\nu - \gamma_\nu \gamma_\mu)$. The invariant $I_1$ with the coupling tensor spinor \eqref{kernel} describes the interactions of the baryons with opposite chiralities. Hence, at high $Q^2$ the corresponding FF is one of the FFs for the processes with quark-helicity flips.

In the PGI model the tensor spinor \eqref{kernel} is a factor of all the couplings permitted by the symmetry. The second differential order introduces two independent invariants $I_2 = i\bar \Psi^{\mu\nu,\omega} \Gamma_{\mu\nu\lambda\sigma} \gamma_\omega V^{\lambda\sigma} N$ and $I'_1 = i\bar \Psi^{\mu\nu} \Gamma_{\mu\nu\lambda\sigma} \gamma_\omega V^{\lambda\sigma} N^{\mu\nu,\omega}$. The latter one, however, is equal to $I_1$ up to a nucleon mass, if the nucleon is on-shell. Thus, the Lagrangian of the second differential order is also defined unambiguously. The invariant $I_2$ relates baryon fields of the same chirality, and at high $Q^2$ the corresponding FF describes the processes conserving quark-helicities.

The third permitted differential order generally contains five independent invariants describing the asymptotic processes with quark-helicity flips. However, the four terms can be reduced to $I_1$ and $I_2$ in the fashion described above.

Finally, the simplest minimally nonlocal PGI Lagrangian of the electromagnetic $N\Delta(1232)$-interactions can be written as follows
\begin{equation}\label{L V}
\begin{aligned}
	\mathscr{L}_{(1)} ={}&
		\sum_V \frac{g_{1}^V}{2M_N^2} \bar \Psi^{\mu\nu}
		\Gamma_{\mu\nu\lambda\sigma} V^{\lambda\sigma} N + \text{H.c.},
	\\
	\mathscr{L}_{(2)}
	={}& \sum_V \frac{i g_{2}^V}{2 M_N^2 M_R^{\hphantom{2}}} \bar \Psi^{\mu\nu,\omega} \Gamma_{\mu\nu\lambda\sigma} \gamma_\omega V^{\lambda\sigma} N + \text{H.c.},
	\\
	\mathscr{L}_{(3)} = {}&
		\sum_V \frac{g_{3}^V}{2M_N^2 M_R^2} \bar \Psi^{\mu\nu,\rho}
		\bigl( \Gamma_{\mu\nu\lambda\rho} g_{\sigma\omega}
		      -\Gamma_{\mu\nu\sigma\rho} g_{\lambda\omega}
		      \\&
		      +\Gamma_{\mu\nu\lambda\omega} g_{\sigma\rho}
		      -\Gamma_{\mu\nu\sigma\omega} g_{\lambda\rho}
		      -\Gamma_{\mu\nu\lambda\sigma} g_{\rho\omega}
		\bigr) V^{\lambda\sigma} N^{,\omega}
	+ \text{H.c.},
\end{aligned}
\end{equation}

The helicity amplitudes for the electroproduction of the $\Delta(1232)$ resonance on the mass shell calculated using the PGI Lagrangian \eqref{L V} are
\begin{align}\label{A32}
	A_{3/2}
	{}&= -\sqrt{N} \Bigl[ \left( Q^2 + \mu M_N \right) F_1
	+ \mu M_R F_2 - \left( Q^2+ \mu M_R \right) F_3 \Bigr],
	\\ \label{A12}
	A_{1/2}
	{}&= -\sqrt{\frac{N}{3}} \Bigl[ \mu M_R F_1
	+ \left( Q^2 + \mu M_N \right) F_2 - \mu M_N F_3 \Bigr],
	\\ \label{S12}
	S_{1/2}
	{}&= -\sqrt{\frac{N}6} Q_+ Q_- \Biggl[ F_1 - F_2
	+ \frac{Q^2+M_R^2+M_N^2}{2M_R^2} F_3 \Biggr],
\end{align}
where $F_f = F_f(Q^2)$, $f = 1,\, 2,\, 3$ are the point and gauge invariant FFs, $F_f(0) = g^\gamma_f$, $\mu=M_R+M_N$, $N=\frac{\pi \alpha Q_-^2}{M_N^5 \left( M_R^2 - M_N^2 \right)}$, $Q_{\pm} = \sqrt{ Q^2 + (M_R \pm M_N)^2}$. Other on-shell observables are expressed in terms of the helicity amplitudes \eqref{A32}--\eqref{S12}. In particular, the magnetic FF \cite{jo-sc} and the ratios $R_\text{EM}$ and $R_\text{SM}$ of the electric and Coulomb quadrupole moments to the magnetic dipole one are written as
\begin{align*}
	G_\text{M}^* &{}=
	- \left[ \frac{M_N^3 (M_R^2-M_N^2)}{2 \pi \alpha \mu^2 Q_-^2} \right]^{1/2}
	 \left( A_{1/2} + \sqrt{3} A_{3/2} \right),
	\\
	R_\text{EM} &{}= \frac{A_{1/2}- A_{3/2}/\sqrt{3}}{A_{1/2}+ \sqrt{3} A_{3/2}},
	\quad
	R_\text{SM} = \frac{\sqrt{2} S_{1/2}}{A_{1/2}+ \sqrt{3} A_{3/2}}.
\end{align*}

It is important to emphasize that the Lagrangian \eqref{L V} is the most general one which is consistent with the considerations of symmetry and locality. All other point and gauge invariant tensor spinor Lagrangian kernels do not affect the kinematic $Q^2$-factors in the helicity amplitudes \eqref{A32}--\eqref{S12}, while might produce nonlocal corrections to the FFs $F_f(Q^2)$
\begin{align}\label{FF series}
	F_f(Q^2) = \sum_{n=0}^{+\infty} F_f^n (Q^2) \left( \frac{Q^2}{4 M_N^2} \right)^n,
	\quad f=1,\,2,\,3.
\end{align}


\section{\label{sec:VMD}FFs as dispersionlike expansions}

Within the vector-meson--dominance (VMD) model \cite{vereshkov:073007}, the FFs $F_f(Q^2)$ are given by dispersionlike expansions
\begin{align}\label{FFs}
	F_f(Q^2) = \sum_{k=1}^{K} \frac{m_k^2\varkappa_{kf}(Q^2)}{m_k^2+Q^2}
	 = \sum_{n=0}^{\infty} \frac{(-1)^n}{Q^{2n}}
	   \sum_{k=1}^{K}{m_k^{2n}\varkappa_{kf}(Q^2)},
\end{align}
with the poles being at the masses $m_k$ of the observed $\rho$-mesons \cite{r2008PDG}. At asymptotically high $Q^2$ pQCD predicts the scaling behavior of the amplitudes to be \cite{ca-86, 1988carlson, 2004idilbi}
\begin{equation}\label{HA scaling}
	A_{3/2} \sim \frac{1}{Q^5 \ell^{n_1}}, \quad
	A_{1/2} \sim \frac{1}{Q^3 \ell^{n_2}}, \quad
	S_{1/2} \sim \frac{1}{Q^3 \ell^{n_3}},
\end{equation}
where $\ell = \ln{(Q^2/\Lambda^2)}$ and $n_2-n_3 \approx 2$ \cite{2004idilbi}.
From Eqs.~\eqref{A32}--\eqref{S12} and \eqref{HA scaling} it follows that the high-$Q^2$ behavior of the FFs is
\begin{align}\label{FF scaling}
	F_1 \sim \frac{1}{Q^8 \ell^{n_1}}, \quad
	F_2 \sim \frac{1}{Q^6 \ell^{n_2}}, \quad
	F_3 \sim \frac{1}{Q^8 \ell^{n_3}}, \quad
\end{align}
and $n_3>n_1$. This implies that the FFs $F_1(Q^2) \sim Q^{-3} A_{3/2}$, $F_2(Q^2) \sim Q^{-3} A_{1/2}$, $F_3(Q^2) \sim Q^{-5} S_{1/2}$ acquire (in the asymptotic domain) the statuses of, respectively, the FF of the processes involving flips of two quark helicities, the non-helicity-flip FF, and the helicity-flip FF.

To assure correct high-$Q^2$ behavior of the dispersionlike expansions of the FFs \eqref{FFs}, we assume the following: (i) The $Q^2$-dependence of the expansion coefficients is independent of the meson index $k$, $\varkappa_{kf}(Q^2) = \varkappa_{kf}(0)/L_f(Q^2)$.
(ii) The interpolation functions $L_f(Q^2)$ are given by $L_f = \left( 1 + b_f \bar\ell + a_f \bar\ell^2 \right)^{n_f/2}$, $\bar\ell = \ln\left( 1+{Q^2}/{\Lambda^2} \right)$, which effectively takes into account the renormalization of the strong coupling constant and the $Q^2$-evolution of the parton distribution functions \cite{1973Brodsky-Farrar,1975Brodsky-Farrar}.
(iii) The parameters of the meson spectrum satisfy the superconvergence relations $\sum_k m_k^{2n} \varkappa_{kf}^{\vphantom2}(0) =0$ for $f=1,\,2,\,3$, $n=1,\,2$ and $f=1,\,3$, $n=3$.


\section{\label{sec:fits}Data analysis and the quasi-scaling}

\begin{figure}
	\center\includegraphics[width=0.85\linewidth]{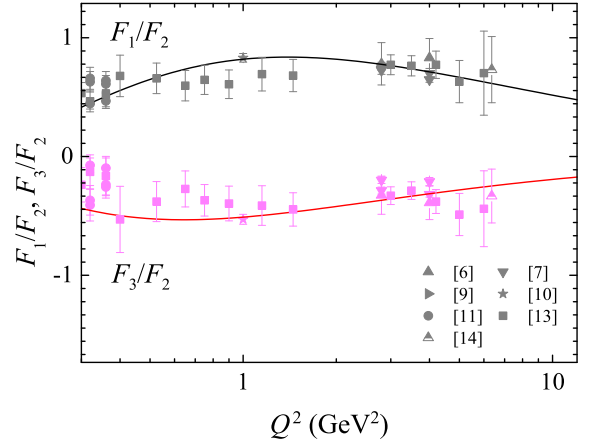}
	\caption{\label{fig:FFRs}FF ratios extracted using Eqs.~\eqref{L V}--\eqref{S12} from the data on the ratios $R_\text{EM}$ and $R_\text{SM}$. The curves are the quasi-scaling fit to the data at $Q^2 = 0.4 - 7\text{ GeV}^2$.}
\end{figure}

The ratios of the PGI FFs $F_{1,3}/F_2$ extracted from the available experimental data \cite{fr-99, ka-01, r2008PDG, sp-05, ke-05, 2007Julia-Diaz, 2008stave, 2009aznauryan, 2009villano} on $R_\text{EM}$ and $R_\text{SM}$ are depicted in Fig.~\ref{fig:FFRs}. Both quantities rise steeply at lower momentum transfers $Q^2 < 1 \text{ GeV}^2$ and decrease slightly at higher $Q^2$. At first glance, it may seem that this peculiar behavior rules out the inverse-square fall-off of the ratios and, consequently, the applicability of the perturbative scaling at these momentum transfers. In fact, it is the behavior that corresponds to the scaling $Q^2$-evolution of the FF ratios
\begin{align}\label{FFR scaling}
	\frac{F_f}{F_2} \propto \frac{\ell^{n_2-n_f}}{Q^2}, \qquad f = 1,\,3.
\end{align}
The aforementioned peculiarities highlight the crucial importance of the logarithmic corrections to the perturbative scaling.

We assume that the FF ratios reach asymptotic scaling \eqref{FFR scaling} at $Q^2 \geqslant 0.4 \text{ GeV}^2$. The results of the fit are shown in Fig.~\ref{fig:FFRs}. The good agreement with the data ($\chi^2/\text{DOF} = 1.03$) testifies that the hypothesis of the quasi-scaling is adequate to describe the $Q^2$-evolution of the ratios $R_\text{EM}$ and $R_\text{SM}$ for the $N\Delta(1232)$-transition.

\begin{figure}
	\center\includegraphics[width=0.81\linewidth]{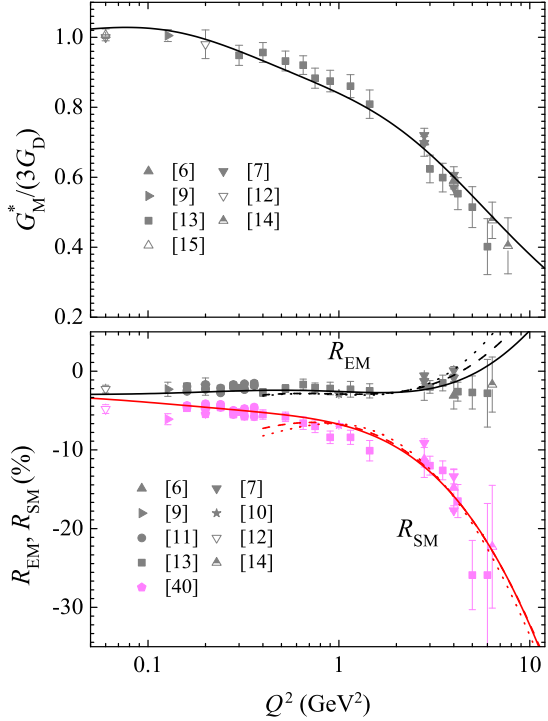}
	\caption{\label{fig:HAs}Magnetic FF divided by three times the dipole FF, $G_\text{D} = (1+Q^2/0.71)^{-2}$, and the ratios $R_\text{EM}$ and $R_\text{SM}$ for the $N\Delta(1232)$ transition. The solid curves are the result of the fit with the dispersionlike expansions \eqref{FFs} of the PGI FFs constrained by the quasi-scaling relations \eqref{FFR scaling}. The dashed and dotted ones are the quasi-scaling fits to the data at $Q^2 > 0.4 \text{ GeV}^2$ with the GI FFs and non-GI FFs, respectively (see text).}
\end{figure}

It is of interest to prove the consistency of the quasi-scaling with the basic principles of effective field theory. An implication of these principles is dispersionlike expansions of the FFs $F_f(Q^2)$, the simplest version being given by the VMD model \eqref{FFs}. To fit the experimental data \cite{fr-99, ka-01, r2008PDG, sp-05, el-06, ke-05, 2007Julia-Diaz, 2008stave, 2009aznauryan, 2009villano, sp-08}, we adjust the parameters of the logarithmic functions $L_f(Q^2)$ and the meson-baryon couplings $\varkappa_{kf}(0)$ for the meson spectrum of the PDG review \cite{r2008PDG}. The ratios of the FFs given by the VMD model \eqref{FFs} are restricted so as not to deviate from the quasi-scaling limit by more than 0.1\% at $Q^2 \geqslant 0.5 \text{ GeV}^2$. The Fig.~\ref{fig:HAs} illustrates that the hypothesis of the quasi-scaling is in satisfactory quantitative agreement with both the experimental data and the dispersionlike expansions \eqref{FFs} of the FFs. The overall quality of the fit by Eqs.~\eqref{A32}--\eqref{FFs} is $\chi^2/\text{DOF} = 1.71$. The parameters of the fit are listed in Tab.~\ref{tab:param}.

To further elucidate the validity of the quasi-scaling, it is necessary to improve the quality of the experimental data base. New experiments should aim at extracting the FF ratios (not simply amplitude ratios), with the emphasis being given to the ratio $F_3/F_2$, which is poorly extracted from the available data. Besides, accurate calculations of the logarithmic corrections to the perturbative scaling become important in the context of the possible quasi-scaling. The fit carried out yields quite large value of the difference $n_2-n_1 = 2.7$. However, the logarithmic corrections ($n_1$) to the amplitude $A_{3/2}$ have not been calculated in pQCD yet.


\section{\label{sec:nPI}Quasi-scaling in conventional models}

Except for the PGI decomposition of the $N\Delta(1232)$-vertex, there exists an infinite number of mathematically consistent alternatives of lower symmetry---gauge but non--point-invariant (GI) models, of which the most popular one is the model discussed in Ref.~\cite{pa-07}. It can be shown straightforwardly that the FFs introduced here are related to the FFs $g_A(Q^2)$, $A=M,\,E,\,C$ of Ref.~\cite{pa-07} by the following relations valid on the mass shell of the RS field,
\begin{align}\label{F1FP}
	F_1(Q^2) &{}= \bar g_M(Q^2) + \bar g_E(Q^2),
	\\
	F_2(Q^2) &{}= \bar g_M(Q^2) - \bar g_E(Q^2) - 2 \bar g_C(Q^2),
	\\ \label{F3FP}
	F_3(Q^2) &{}= -2 \bar g_C(Q^2),
\end{align}
where $\bar g_A (Q^2)= \sqrt{\frac32} \frac{\mu M_N}{2 Q^2_+} g_A (Q^2)$.

From pQCD predictions \eqref{HA scaling} and Eqs.~\eqref{A32}--\eqref{S12}, \eqref{F1FP}--\eqref{F3FP} it is seen readily that at high $Q^2$ the FFs $\bar g_A(Q^2)$ cannot be made to scale as different helicity amplitudes. Hence, to comply with pQCD constraints, the FFs $\bar g_M(Q^2)$ and $\bar g_E(Q^2)$ should mix the contributions of different-scale asymptotic quark-gluon subprocesses,
\begin{align*}\label{}
	\bar g_M(Q^2) &{}= \frac{L_1(Q^2)}{Q^6} + \frac{L_2(Q^2)}{Q^8},
	\\
	\bar g_E(Q^2) &{}= -\frac{L_1(Q^2)}{Q^6} + \frac{L_3(Q^2)}{Q^8},
	\qquad
	\bar g_C(Q^2) = \frac{L_4(Q^2)}{Q^8},
\end{align*}
where $L_i(Q^2)$ are slowly varying functions. This obscures the quasi-scaling behavior of the FFs in the model for two reasons. Firstly, from the above asymptotic relations it is clear that the ratios of the FFs $g_A(Q^2)$ do not exhibit the scaling behavior analogous to \eqref{FFR scaling}, but only some linear combinations thereof do---for instance, the combinations \eqref{F1FP}--\eqref{F3FP}. Secondly, in the frame of this model any linear combination of the FFs (e.g., $g_M-g_E$) can play the role of the non--helicity-flip FF at high $Q^2$ as well as any linear combination of the FF $\bar g_C(Q^2)$ and the sum $\bar g_M(Q^2)+\bar g_E(Q^2)$ can play the role of the FF for the processes with two quark-helicity flips. Thus, in the general case, the condition of gauge invariance does not give a definite interpretation to the FFs in terms of the quark-gluon dynamics. In this sense the requirement of the point invariance is a symmetry condition that excludes such an ambiguity by proving us with a set of FFs that manifestly correspond to different quark-gluon subprocesses in the asymptotic domain.

The simplest way to impose the pQCD constraints on the FFs $\bar g_A(Q^2)$ is to suppose that $L_1(Q^2)\sim \ell^{-n_2}$, $L_2(Q^2) = L_3(Q^2) \sim \ell^{-n_3}$, and $L_4(Q^2) \sim \ell^{-n_3}$. The hypothesis of the scaling of the ratios $(g_M+g_E)/(g_M-g_E)$, $g_C/(g_M-g_E)$ agrees with the data \cite{fr-99, ka-01, r2008PDG, sp-05, ke-05, 2007Julia-Diaz, 2008stave, 2009aznauryan, 2009villano} at $Q^2 \geqslant 0.4 \text{ GeV}^2$ with $\chi^2/\text{DOF} = 1.57$ (cf. $\chi^2/\text{DOF} = 1.07$ for the PGI FFs). It suggests that the quasi-scaling behavior is a quite general property of the gauge invariant FFs, although the ambiguity in the definitions of the FFs can obscure the quasi-scaling. The fit results are depicted in Fig.~\ref{fig:HAs} (bottom panel).

It is interesting that the quasi-scaling of the another popular (non--point- and non--gauge-invariant) choice of the FFs \cite{Devenish1976, vereshkov:073007} is in much poorer agreement with the data, $\chi^2/\text{DOF} = 2.58$ (The curves are shown in Fig.~\ref{fig:HAs}, the bottom panel.) Perhaps, it emphasizes the importance of the intrinsic spin-$3/2$ symmetries for constructing the $N\Delta(1232)$-interactions. 


\section{\label{sec:conclusion}Conclusion}

We have constructed an effective-field model of the electromagnetic $N\Delta(1232)$-transition. The interaction Lagrangian of the model possesses the gauge and point invariance of the RS field. This ensures mathematical coherence of the theory and fixes the pre-FF $Q^2$-polynomials in the observables at the peak of the resonance.

Within the developed model, a class of observables---the ratios of the helicity-flip FFs to the non-helicity-flip one---exhibits asymptotic scaling behavior at momentum transfers as low as $0.4 \text{ GeV}^2$. Put in other words, this gives a new interpretation of the available experimental data on quadrupole transitions---though it is widely believed that the applicability of the scaling is ruled out by the data at $Q^2 < 7 \text{ GeV}^2$ \cite{2009aznauryan}, the data available validate the hypothesis of the quasi-scaling of the $N\Delta$-transition FFs. Although our analysis is phenomenological in its character, it poses questions to the underlying theory of the strong interactions---while the scaling of the FFs is well understood as a consequence of the asymptotic freedom, the dynamics leading to the quasi-scaling in the nonperturbative domain of QCD is still to be established both qualitatively and quantitatively.

Finally, our results make it important to point out that the approach to pQCD of the $N\Delta(1232)$ transition should be studied in terms of the vertex FFs $F_f(Q^2)$ or similar quantities. Such observables as helicity amplitudes, ratios thereof, and others contain kinematic $Q^2$-dependent factors that can ``disguise'' the transition to asymptotic QCD predictions. As has been shown in Section \ref{sec:fits}, this is what appears to happen with the ratio $R_\text{EM}(Q^2, F_1/F_2, F_3/F_2)$ at $Q^2 < 7 \text{ GeV}^2$ that is seemingly well out of the domain of perturbative evolution, but, in fact, agrees well with pQCD if the scaling behavior of the point and gauge invariant FF ratios is assumed. On the other hand, to take into account kinematic factors postulated in effective field theory may be essential to achieve quantitative agreement with the experimental data. This can be seen if we contrast our fits with those of Ref. \cite{1998carlson}. In Ref. \cite{1998carlson} it is suggested that the helicity amplitudes for $N\Delta(1232)$ transition can exhibit perturbative scaling at moderate momentum transfers. However, although this hypothesis may explain qualitatively the observed $Q^2$-evolution of the ratio $R_\text{EM}$, it still disagrees with the data quite significantly, which is pointed out in Ref. \cite{fr-99}. As we have seen, this result can be amended if we assume the scaling of the FF ratios (not helicity amplitudes) and correctly take into account kinematic $Q^2$-dependence of the amplitudes predicted by effective field theory.

New experimental data from ongoing and proposed experiments will test the proposals of this Letter and are eagerly awaited. A more detailed discussion on the issues of the Letter along with the construction of the Lagrangian \eqref{L V} and more thorough data analyses will be given in a separate paper.

\begin{table}
\caption{\label{tab:param}Fit parameters and PDG meson masses \cite{r2008PDG}.}
\small
\begin{tabular}{cD{.}{.}{-1}ccD{.}{.}{-1}}
$\varkappa_{51}(0)$ & 0.0001 &\qquad\quad& $a_1$ & 0.0090 \\
$\varkappa_{42}(0)$ & -0.8642 && $b_1$ & -0.1413 \\
$\varkappa_{52}(0)$ & 0.8683 && $a_2$ & 0.2268 \\
$\varkappa_{53}(0)$ & -0.0341 && $b_2$ & -0.1339 \\
$\sum_{k=1}^5 \varkappa_{k1}(0)$ & 0.42026 && $a_3$ & 0.9416 \\
$\sum_{k=1}^5 \varkappa_{k2}(0)$ & 0.75559 && $b_3$ & -0.0377 \\
$\sum_{k=1}^5 \varkappa_{k3}(0)$ & -0.31849 && $m_1$ & 0.77549 \\
$n_1$ & 0.2276 && $m_2$ & 1.465 \\
$n_2$ & 3 && $m_3$ & 1.72 \\
$n_3$ & 1 && $m_4$ & 1.88 \\
$\Lambda$ & 0.2950 && $m_5$ & 2.149
\end{tabular}
\end{table}

\end{document}